\documentclass[useAMS,usenatbib,usegraphics]{mn2e}

\usepackage{graphicx,epsfig}  

\def\ut#1{\mathop{\vtop{\ialign{##\crcr
     $\hfil\displaystyle{#1}\hfil$\crcr\noalign
     {\kern1pt\nointerlineskip}\hbox{$\hfil\sim\hfil$}\crcr
     \noalign{\kern1pt}}}}}

\def\undersymbol#1#2{\mathop{\vtop{\ialign{##\crcr
     $\hfil\displaystyle{#2}\hfil$\crcr\noalign
     {\kern1pt\nointerlineskip}\hbox{$\hfil#1\hfil$}\crcr
     \noalign{\kern1pt}}}}}

\title[Pixel-lensing as a way to detect extrasolar planets in M31]
{Pixel-lensing as a way to detect extrasolar planets in M31}
\author[G. Ingrosso et al.]
{G. Ingrosso,$^1$\thanks{E-mail:
ingrosso@le.infn.it}
 S. Calchi Novati,$^2$
 F. De Paolis,$^1$
\newauthor Ph. Jetzer,$^3$
A.A. Nucita$^4$ and
A.F. Zakharov$^{5,6}$\\
$^1$  Dipartimento di Fisica, Universit\`a del Salento and
         {\it INFN} Sezione di Lecce, CP 193, I-73100 Lecce, Italy \\
$^2$         Dipartimento di Fisica, Universit\`{a} di Salerno,
       I-84081 Baronissi (SA) and {\it INFN} Sezione di Napoli, Italy\\
$^3$       Institute for Theoretical Physics,
           University of  Z\"{u}rich, Winterthurerstrasse 190,
           CH-8057 Z\"{u}rich, Switzerland\\
$^4$        XMM-Newton Science Operations Centre, ESAC, ESA,
            PO Box 50727, 28080 Madrid, Spain \\
$^5$       Institute of Theoretical and Experimental Physics,
            B. Cheremushkinskaya 25, 117259 Moscow,  Russia \\
$^6$      Bogoliubov Laboratory of Theoretical Physics, Joint Institute for Nuclear Research, 141980 Dubna, Russia}
\begin{document}

\date{Accepted xxx; Received xxx;
in original form xxx}

\pagerange{\pageref{firstpage}--\pageref{lastpage}} \pubyear{2008}

\maketitle

\label{firstpage}

\begin{abstract}

We study the possibility to detect extrasolar planets in M31
through pixel-lensing observations. Using a Monte Carlo approach, we select
the physical parameters of the binary lens system, a star
hosting a planet, and we calculate the pixel-lensing light curve
taking into account the finite source effects.
Indeed, their inclusion is crucial since the sources
in M31 microlensing events are mainly giant stars.
Light curves with detectable planetary features are selected by
looking for significant deviations from the corresponding
Paczy\'{n}ski shapes.
We find that the time scale of planetary deviations in  light curves
increase (up to 3-4 days) as the source size increases.
This means that only few exposures per
day,  depending also on the required accuracy,
may be sufficient to reveal in the light curve
a planetary companion.
Although the mean planet mass for the selected events is about
$2~M_{\rm {Jupiter}}$, even small mass planets ($M_{\rm P} < 20~M_{\oplus}$)
can cause significant deviations,
at least in the observations with large telescopes.
However, even in the former case, the probability to find
detectable planetary features in
pixel-lensing light curves is at most a few percent of the detectable
events, and therefore many events have to be collected in order to
detect an extrasolar planet in M31.
Our analysis also supports the claim
that the anomaly found in the candidate event
PA-99-N2 towards M31 can be explained by a companion object orbiting
the lens star.

\end{abstract}


\section{Introduction}

In the last years, it has become clear that gravitational
microlensing, initially developed to search for MACHOs in our
Galactic halo and near the Galactic disk
\citep{pacz86,macho93,
Paczynsky_96, Roulet_97,Roulet_02,Zakharov_Sazhin_UFN98} can be used
to infer the presence of extrasolar planets orbiting around lens
stars (see the review by
\citealt{perryman2000,Perryman2005,bennett09}).

As shown by \cite{maopacz91} the planet presence effect
on the light curve in a microlensing event
towards the Galactic bulge  is generally a
short duration perturbation to the standard microlensing curve.
These deviations last from a few hours to some days (depending on
the planet mass) and can occur relatively frequently, even for
rather small mass planets. Indeed, the microlensing technique is
sensitive to planets in a rather large range of masses, from
Jupiter-like planets down to Earth-like ones \citep{bennett96}.

\cite{gould92} pointed out that there is a significant probability
to detect planets around stars in the Galactic disk that act as
microlenses by magnifying the light of observed stars in the
Galactic bulge. Until now, the detection of eight extrasolar planets
has been reported by using the microlensing technique
\citep{bond04,udalski05,beaulieu06,gould06,gaudi08,bennett09}.
We remind that the masses of three of them
($\simeq$ 3, 5 and 13 $M_{\oplus}$)
are at the lower bound of the detected planetary mass range.
Indeed, more than 300 extrasolar planets
discovered until now by radial velocity, transit and direct imaging
methods are biased towards large mass (Jupiter-like) planets \citep{ida}.
However, radial velocity searches by ground based experiments
have now provided extrasolar planets with
$M_{\rm min} = 2~M_{\oplus}$ \citep{Mayor},
whereas space based observations are expected to detect
many Earth-mass planets (with Kepler satellite
\footnote{http://www.nasa.gov/mission\_pages/kepler/overview/index.html})
and many Earth-size planets (with COROT spececraft
\footnote{http://smsc.cnes.fr/COROT/index.html}).

A further advantage of the microlensing is that
it works better for large distance of the source star,
since the optical depth increases by increasing the distance,
as one can already see from the \cite{einstein36} approach.
This gives the opportunity to detect planetary systems
at distances much larger with respect to those
accessible by the other tecniques and even in other galaxies such as
M31 (see, e.g., \citealt{covone,baltzgondolo2001}).
In this case, however, the source stars are not resolved by ground based
telescopes -  a situation referred to as ``pixel-lensing''
\citep{crotts92,agape93,gould96} - and only bright sources
(i.e. giant stars with large radii), sufficiently magnified,
can give rise to detectable microlensing events \citep{agape97}.
This implies that finite size effects, leading to smaller planetary deviations
in pixel-lensing light-curves with respect to microlensing towards the galactic
bulge, cannot be neglected (see, e.g., \citealt{Riffeser08}).
Usually, highly magnified events arise when the source and lens
stars align very closely. In this case there is the largest chance
of observing the perturbations in the light curves induced by planets
\citep{griest98}.
This is particularly true for large mass planets, for which the planetary
signals are not strongly suppressed by finite size effects,
whereas for low mass planets, the planetary signals may remain detectable
during other phases of the event \citep{bennett09}.

Until now, only about a dozen microlensing events have been observed
towards M31 by the POINT-AGAPE \citep{novati05} and MEGA
collaborations \citep{dejong06}. Only in one case a deviation
from the standard Paczy\'{n}ski shape has been observed
and attributed to a
secondary component orbiting the lens star \citep{an04}.
However, new observational campaigns towards M31 have been
undertaken \citep{kerinsetal06,novati07,novati09}
and hopefully a few planets might be  detected  in the future,
providing a better statistics on the masses and orbital radii of
extrasolar planets. It is in fact expected, and supported by
observations and numerical simulations, that almost any star has at
least a planet orbiting around it
(see, e.g., \citealt{lineweaver}).
In other words, as also suggested
by  \cite{baltzgondolo2001}, the rate of single lens events towards
M31 may suffer of a strong contamination of binary lensing events,
most of which are expected to be due to extrasolar planets.

Therefore, it is  important to address the question of how to extract
information about planetary lensing events, occurring in M31, from the
observed microlensing light curves.
Since planetary perturbations last from hours to a few days,
a monitoring program with suitable
sampling must be realised, in order to avoid missing these perturbations.
The feasibility of such research program has been already explored
by \cite{chungetal06} and \cite{kimetal07}. They have considered
the possibility to detect planets in M31 bulge by using the
observations taken from the Angstrom collaboration
\citep{kerinsetal06} with a global network of 2 m class telescopes
and a monitoring frequency of about five observations per day. The
analysis for planet detection, however, has been performed by using
a fixed configuration of the underlying Paczy\'{n}ski light curve.

In the present work, instead, by using a Monte Carlo (MC) approach
\citep{ingrosso05,ingrosso06,ingrosso07} we explore the possibility of
detecting extrasolar planets in pixel-lensing observations towards M31,
by considering the multi-dimensional space of parameters for
both lensing and planetary systems.
Taking into account the finite source effects and the limb darkening
and using the residual method we can select the simulated light curves that
show significant deviations with respect to a Paczy\'{n}ski like
light curve, modified by finite source effects.
The advantage of the Monte Carlo approach is that of allowing
us a complete characterization of the sample of microlensing events
for which the planetary deviations are more likely to be detected.

The paper is structured as follows. In Section 2, we give the basics
of binary-lensing events. In Section 3 we discuss our
MC simulations for planetary detection in M31.
In section 4 we present our main results and in Section 5
we address the conclusions.

\section[]{Binary-lensing events}

\subsection{Generalities}

If a source star is gravitationally lensed
by a binary lens, the equation of lens mapping from the lens plane
to the source plane can be expressed in complex notation
\citep{witt90,witt95}
\begin{equation}
\xi(\zeta,
\eta)=z-\sum_{j=1}^{2}\frac{m_j/M}{\bar{z}-\bar{z}_{L,j}}~,
\label{mao}
\end{equation}
where $\xi=\zeta+i\eta$ and $z=x+iy$ are the source and the image
positions, $\bar{z}$ is the complex conjugate of $z$, $m_1$, $m_2$,
$z_{L,1}$ and $z_{L,2}$ are the the masses and the positions of the
two lenses, respectively.  Here and in the following, all the lengths
(angular separations) are normalized to
the radius $R_{\mathrm E}$ (angle $\theta_{\rm E}$)
of the Einstein ring  which are related to
the physical parameter of the lens by
\begin{equation}
R_{\mathrm E}  = \left[ \left( \frac {4GM} {c^2} \right)
\frac {D_{\rm L} (D_{\rm S}-D_{\rm L})} {D_{\rm S}} \right]^{1/2}~~~~~
{\rm and}~~~\theta_{\rm E}= \frac{R_{\rm E}}{D_{\rm L}}~,
\end{equation}
where $M=m_1+m_2$ is the total mass of the binary system,
$D_{\rm L}$ and $D_{\rm S}$ are the distances to the lens and to the
source, respectively.
Under the condition $m_1 > m_2$, we define
the mass ratio parameter $q = m_2/m_1$. In addition,
we assume that the two masses of the
binary system are located on the real axis,
with the centre of mass
in the origin. Let us denote with
$d$ the angular separation between the
two objects in units of $\theta_{\rm E}$.

To determine the image position and magnification, one has to take
the complex conjugate of equation (\ref{mao})
and substitute the expression for $\bar{z}$ back in it,
obtaining  a fifth-order polynomial in $z$,
i. e. $p(z)=\sum_{i=1}^{5}c_iz^i=0$
(with coefficients $c_i$ depending on $M$, $d$, and $q$),
whose solutions give the image positions.
Due to lensing, the source star image splits into several
fragments up to a total number $N_I$. Since the lensing process
conserves the source brightness and thus the magnification
of each image, the total magnification corresponds to the sum over
all images \citep{witt95}, i. e.
\begin{equation}
A_{\rm P} = \sum_i^{N_I}\left(\frac{p(z_i)}{\det J}\right)~,
\end{equation}
where the determinant of the Jacobian is
\begin{equation}
\det J = 1-\frac{\partial \xi}
{\partial \overline{z}}\frac{\overline{\partial \xi}}{\partial \overline{z}}~.
\end{equation}

A planetary lens system is characterized by the condition that the
planet mass $M_{\rm P}=m_2$ is much smaller with respect to the host star mass
$M_{\rm L}=m_1$. In this case, the planet only induces a perturbation on the
underlying Paczy\'{n}ski curve of the primary lens. Planet perturbations
occur when the source star crosses and/or passes near caustics,
which are the set of source positions on the $(\zeta, \eta)$
plane  at which the magnification is infinite (i.e. those
corresponding to $\det J=0$) in the idealized case of a point
source. Clearly, for realistic sources of finite size the magnitude
gets still quite large, but finite  \citep{witt94}.
Caustics form a single or multiple sets of closed and concave curves
(fold caustics)
which meet in cusp points
\citep{SEF,sch92,Zakharov_95}.
The location of the
planet perturbations depends on the position of the caustics and the
source trajectory.

There have been several attempts to determine caustic positions and shapes
by using analytic methods and treating the planet induced deviations as a
perturbation \citep{gaudi97,dominik99,bozza99}.
For the planetary case, there exists two sets of caustics: ``central'' and
``planetary''. The single, central caustic is located on the star-to-planet
axis, close to the host star.  For a wide range of parameters the caustic has a
diamond shape and can be described by parametric equations
(as it was shown by
\citealt{Zakharov_Sazhin_97a,Zakharov_Sazhin_97b}, central astroid
caustics arise if the \citealt{Chang84a,Chang84b} model is used).
Planetary caustics are located away from the host star, at distance
$\simeq (d^2-1)/d$ from the primary lens position.
There is one planetary caustic (with a diamond shape)
on the star-to-planet axis, on the planet side, when $d>1$
and two sets of caustics, off the axis, (with triangular
shape) on the star side when $d<1$.
The dimensions of both central and planetary caustics
increase by increasing the mass ratio $q$
\citep{Zakharov_Sazhin_UFN98,bozza99,chungetal05,Han_Gaudi_08}.
Moreover, for a given $q$ value, the caustic sizes are maximized
when the planet is inside the so called ``lensing zone'', which
is defined (with some arbitrariness) as the range of star-to-planet separation
$0.6  \ut <  d \ut < 1.6$ \citep{gould92,griest98}.
The time duration scale of the perturbations induced by a planet and the
probability of their detection are proportional to the caustic size,
at least when this region is large enough so that the planetary signals are not
suppressed by the finite source effects
\citep{maopacz91,bolatto,gould92}.

\subsection{Finite source approximation}

Since in pixel-lensing towards M31 the bulk of the source stars
are red giants (see Section 3), one has to take into account the
source finiteness. This leads to smaller planetary deviations in pixel-lensing
light curves with respect to microlensing towards the galactic bulge,
for which the point-like source approximation is acceptable.
For finite source effects with limb darkening
the magnification has to be numerically evaluated
(see, e.g., \citealt{SEF,Bogdanov95a,Dominik05} and references therein)
\begin{equation}
\langle A_{\rm P}(t) \rangle = \frac {\int_0^{2\pi} d\theta \int_0^{\rho}
A_{\rm P}(\tilde{\theta},\tilde{\rho};t)I(\tilde{\rho})
\tilde{\rho}d\tilde{\rho}} { 2 \pi \int_0^{\rho} I(\tilde{\rho})
\tilde{\rho}d\tilde{\rho}} ~, \label{finite1}
\end{equation}
 where $\rho = { \theta_{\rm S} }/ { \theta_{\rm E} }$
is the normalized angular size of the source
($\theta_{\rm S} = R_{\rm S}/D_{\rm S}$, $R_{\rm S}$ being the source radius),
and $I(\tilde{\rho})$ is the intensity
profile of the source including limb darkening,
for which we use the \cite{claret}
approximation
\begin{equation}
 I(\tilde{\rho}) =
1-a_1(1-\mu^{1/2})-a_2(1-\mu)-a_3(1-\mu^{3/2})-a_4(1-\mu^{2})~,
\end{equation}
with $\mu = \tilde{\rho}/\rho$ and the coefficients in the $R$-band
\footnote{http://webviz.u-strasbg.fr/viz-bin/VizieR-source=J/A+A/363/1081.}
are $a_1=0.8282$, $a_2=-0.9866$, $a_3=1.6801$ and $a_4=-0.6604$
(for red giant stars).

Moreover, since during caustic crossing the  magnification  could have
strong changes and (at least for small mass planets and/or realistic
source sizes) typical time scale for crossing could be
comparable with the exposure time $t_{\rm exp}$ (needed to have a
reasonable signal-to-noise level) we take the average  magnification
of equation (\ref{finite1}) in the interval
$(t-t_{\rm exp}/2,t+t_{\rm exp}/2)$.

Finite size source effects can be relevant for two reasons. First, the
relationship between the dimensionless radius $\rho$ and the impact parameter
$u_0$ determines if the finite size effects are important or not
for the main microlensing light curve.
This occurs in the events with $\rho/u_0>1$ or $\rho/u_0<1$, respectively.
Second, finite size effects may be important for the planetary deviations
even if they are not relevant for microlensing without planets.
Indeed, $\rho$ is to be compared not only with $u_0$,
but also with the caustic size $\Delta$.
In particular, whenever $\rho/u_0>1$, it results that
$\rho$ is typically much larger than $\Delta$
(at least for small enough mass planets). In this case, smoothed planetary
deviations are produced in the light curves, since
the planetary  magnification
has to be averaged on the source area.
In a similar way, depending on the lens system geometry and proper motion,
whenever the ratio $\rho/u_0<1$, stronger and temporally localized
planetary deviations are produced in the light curves, since the caustic
region results to be a non negligible fraction of the source area.
Within the following analysis for the detection of planetary deviations we are
going to identify two classes (I and II) of events, depending on the ratio
$\rho/u_0>1$ and $\rho/u_0<1$, respectively
\footnote{
We mention the classification of the planetary perturbations
by \cite{covone}, which distinguishes two main types of anomalies in the light
curves, namely the events affected by the central caustic (type I), and the
ones affected by one of the planetary caustics (type II).
In our analysis we do not attempt to characterize the
planet deviations as due to the intersection of central and/or planetary
caustics, but we look at the shape of the induced planetary features on the
light curves. A classification of the events based on the caustic crossing
will be considered in a forthcoming paper.}.

\section[]{Monte Carlo simulation}

\subsection{Light curve generation}

In the present analysis we assume that the lens is a binary system
constituted by a star and a planet companion
\footnote{Based on the recent  detections of multiple planets
\citep{gaudi08}, one can expect that this assumption is rather conservative.}.
Our aim is to evaluate the probability to detect the presence of
planets in M31 through Earth-based pixel-lensing observations
with telescopes of different diameters. These telescopes
could be initiated to observe towards a microlensing event
candidate, so making a high cadence observations of an ongoing
microlensing event.
As reference values, we adopt a CCD pixel field
of view of  0.2 arcsec, a typical seeing value of 1 arcsec and
an average background luminosity at telescope site of $\simeq 21$
mag arcsec$^{-2}$ in $R$-band. To have a good S/N ratio
we consider in the MC analysis exposure times $t_{\rm exp}$ of
30~minutes.
 Moreover, we assume a regular sampling neglecting any
loss of coverage due to bad  weather conditions.

In order to take into account the spatial variation of the
background level we select four directions (named ${\bf A,~ B,~ C,~ D}$)
at increasing distances from the M31
centre. Assuming a coordinate system with origin in the M31 centre
and axes along the north-south and east-west directions, the
coordinates of the selected directions are the following:
 $\bf A$ (--6,0) arcmin,
 $\bf B$ (--9,0) arcmin,
 $\bf C$ (--12,0) arcmin,
 $\bf D$ (--21,--6) arcmin.
In the direction $\bf A$ the microlensing is primarily due to
self-lensing events by stars in the M31 bulge and disk, whereas
towards the external directions the contribution to microlensing due
to lenses belonging to the M31 halo becomes larger. Our investigation
of the $\bf D$ direction is motivated by the  detection of the
anomaly in the pixel-lensing event  PA-99-N2 \citep{an04}.

As for the generation of the trial microlensing light curves we
closely follow the approach outlined by \cite{kerins01}. The adopted
M31 astrophysical model was described by \cite{ingrosso06}.
Once the event location has been selected, for
any lens and source population lying
along the line of sight, we
use a MC approach to select the
physical parameters of the systems:
source magnitude, primary lens mass,
source and lens distances, effective transverse velocity of source and lens,
impact parameter of the lens.

The luminosity of star sources, mainly red giants in the interval of
absolute magnitude $(-4,~2.4)$,
and the corresponding radii are drawn from a sample of stars
generated by a synthetic color-magnitude diagram computation algorithm
\footnote{http://iac-star.iac.es/iac-star.} described by
\cite{aparicio} based on the stellar evolution library
\citep{bertelli94} and the bolometric correction database
\citep{girardi02}.

As next, we have to select the mass $M_{\rm P}$ and the (projected)
orbital distance
$d_{\rm P}$ of the extrasolar planet. Most of the hundreds
of extrasolar planets discovered up to now
(see the web site http://exoplanet.eu)
have typically very large masses and orbit at
small distances around their parent stars \citep{Udry07}. This
appears to be a result of observational biases \citep{ida} since
most of the planets have been  detected by radial velocity and
transit techniques that are most sensitive to massive and close
planets. Direct imaging and microlensing techniques contribute only
a minor fraction of the detected events. Indeed, available theoretical
and numerical analysis  show that most extrasolar planets are expected
to have relatively smaller masses. Furthermore, the (projected)
orbital distance from
their hosting stars is expected in the range $\simeq 0.04 - 100$ AU
(see, e.g., \citealt{tremaine,ida}). In the present paper we assume that the
distribution of $M_{\rm P}$ and orbital period $P$,
for $M_{\rm P} < 10 M_{\rm {Jupiter}}$,
is given by the simple analytical expression
\citep{tremaine}
\begin{equation}
dn(M_{\rm P},P) = C~ M_{\rm P}^{-\alpha}~ P^{-\beta}~
\left(\frac {dM_{\rm P}}{M_{\rm P}}\right) ~ \left( \frac {dP}{P}\right)~~,
\label{dTT}
\end{equation}
with $\alpha=0.11$ and $\beta=-0.27$.
This relation is obtained by investigating the distribution of masses
and orbital periods of 72 extrasolar planets, taking into account
the selection effects caused by the limited velocity precision and
duration of existing surveys.
We note that in the analysis leading to the above distribution,
it was assumed that the stars in the survey are of solar type, and therefore
any dependence (as implied by recent extrasolar planet observations) of the
planet mass on the parent star mass and metallicity has been neglected.
Taking that into account, one would
certainly replace equation (\ref{dTT}) with a different one, and therefore the
results presented in Section 4 for the detectable planet rate would change.
For example, a steeper planet mass distribution
(as that found for all Doppler-detected planets by \cite{johnson}
with $\alpha = 0.4$) implies a smaller (about a quarter)
overall planet detection rate, as a consequence
of the decrease of the mean planet mass.
More importantly, a dependence of the planet mass distribution
on the parent star mass would introduce a dependence of the planet detection
rate on the lens population (bulge or disk stars)
that could be recognized, provided a sufficient event statistics
towards different lines of sight would be available.
In equation (\ref{dTT}), the upper limit of the planetary mass
is set at
$M_{\rm P}= 10~M_{\rm {Jupiter}}$. This roughly corresponds to the
usually assumed lower mass limit for brown dwarfs. Indeed, in the range
$10 - 20~M_{\rm {Jupiter}}$ the two populations overlap.
Moreover, in the simulation
we select a lower planetary mass limit of $0.1~M_{\oplus}$.
Once the masses of the binary components and the planet period have
been selected, the binary separation $d_{\rm P}$ is obtained by assuming a
circular motion of the planet.

As a parameter in our MC analysis we introduce the number
$N_{\rm im}$ of images per day. We take $N_{\rm im}$ in the range 2 -- 12 day$^{-1}$,
the latter value corresponding to a sampling time
of two hours.
For all selected values of $N_{\rm im}$, the corresponding binary
light curve  at any time is given by
\begin{equation}
S_{\rm P}(t) = f_{\rm bl} + f_0 \left( \langle A_{\rm P}(t) \rangle -1\right)~,
\label{8888}
\end{equation}
where $f_{\rm bl}$ is the  background signal
from the galaxy and the sky,
$f_0$ is the unamplified source star flux
and $ \langle A_{\rm P}(t) \rangle $ the time varying  magnification,
that takes into account both the source finiteness and the
motion of the lens-source-observer system during
the exposure time $t_{\rm exp}$.
To mimic superpixel photometry \citep{agape97}
used in a real observational campaign we evaluate the star and the
background flux within a $n$-pixel square ``superpixel'', whose size $n$ is
determined to cover most of the average seeing disk.
We recall that we consider the pixel-lensing regime where the
noise is dominated by the (line of sight dependent) background noise
\citep{kerins01}. Accordingly, we add to $S_{\rm P}(t)$ a Gaussian noise.

\subsection{Microlensing event selection}

As a first step, within the MC simulation, we have to test whether
the flux variation due to the microlensing event is
significant with respect to
the background noise $\sigma (x,y)$, where $(x,y)$ identifies the
line-of-sight. To asses the detection of a flux variation we
evaluate its statistical significance testing whenever and to what
extent at least three consecutive points exceed the baseline level
by 3$\sigma$, following the analysis described by \cite{novati02}.
We remark that the condition on the variation significance is the
only one used at this stage. In the following we refer to events
that show a significant flux variations as to ``detectable'' events.

\subsection{Planet detection}

The expected signature of an extrasolar planet orbiting the lens star is
the presence of perturbations with respect to the corresponding smooth
Paczy\'{n}ski light curve. Therefore, we look for a selection criterion based
on the analysis of the significance of such deviations.
To this purpose, given the wide range of the microlensing parameters and
the corresponding planetary deviations, we  consider two indicators for
which we select (by the direct survey of many light curves) threshold values.
They are the mean deviation (in units of $\sigma$) of the planetary light
curve from that of a single lens event, and the maximum value of the time
dependent relative planetary  magnification  (in units of the expected
Paczy\'{n}ski value).

At first, we fit the light curve in equation (\ref{8888})
with a  Paczy\'{n}ski-like law
modified to take into account finite source effects
and determine the best fit parameters.
The latter  are the baseline flux $f_{\rm bl}^0$,
the maximum  magnification  time $t_0^0$,
the unamplified star flux      $f_0^0$,
the Einstein time              $t_\mathrm{E}^0$,
the dimensionless impact parameter  $u_0^0$ and
the dimensionless, projected star radius $\rho^0$.
Accordingly, the time dependent  flux $S^0(t)$
due to a single lens event is given by
\begin{equation}
S^0(t) = f_{\rm bl}^0+f_0^0 \left( \langle A^0(t) \rangle -1\right)~,
\end{equation}
where the magnification \citep{einstein36,pacz86}
\begin{equation}
A^0(t) = \frac {u^0(t)^2+2} {u^0(t) \sqrt{u^0(t)^2+4}}
\label{apacz}
\end{equation}
is given in terms of the time varying (normalized)
lens angular distance to the source
$u^0(t)$
\begin{equation}
u^0(t) = \sqrt { [u^0_0]^2 + [(t-t^0_0)/t^0_{\rm E}]^2} ~,
\end{equation}
and $ \langle A^0(t) \rangle $ is the analogous of equation (\ref{finite1}),
in this case evaluated
by using the analytical approximation given by \cite{witt94}.
Then, we can evaluate the time dependent variable
\begin{equation}
\chi^2(t) = (S_{\rm P}(t)-S^0(t))^2/\sigma^2(t)~,
\end{equation}
and the residual to the single lens fit
\footnote{The analysis of residuals is a well known technique widely
applied to search for deviations with respect to a null hypothesis.}
\begin{equation}
\chi_{\rm r}^2(t) = \left(1- \chi^2(t)\right)^2~,
\end{equation}
where $S_{\rm P}(t)$ is the light curve including the planet perturbations,
      $S^0(t)$ the Paczy\'{n}ski fit
as above and $\sigma(t)$ is evaluated according to \cite{kerins01}.
Therefore, we can consider large values of $\chi_{\rm r}^2(t)$
as a significant indicator of the presence of
detectable planetary deviations in the light curves.
Actually, we use the sum of the residuals along the whole light curve,
namely $\chi_{\rm r}^2 = \sum_i \chi_{\rm r}^2(t_i)/N_{\rm tot}$,
as a first quantitative measure of the statistical significance of the
planetary signals in the ongoing microlensing event.
Here $t_{\rm i} = t_0 + [(t_{\rm f}-t_0)/N_{\rm tot}]~i$, where
$t_0$ ($t_{\rm f}$) is the initial (final) instant and $N_{\rm tot}$
the total number of  points.
By the direct survey of many light curves we select a threshold value
\footnote{
We find that the residuals $\chi_{\rm r}$ follow a Gaussian distribution
(with mean value $\simeq 1.4$ and standard deviation $\simeq 0.3$)
in the case of light curves generated as single lens events and subsequently
fitted with a Paczy\'{n}ski law.}
$\chi_{\rm r{~\rm th}} = 4$.
We further require a minimal number of points
$N_{\rm good}$, even not consecutive, which deviate significantly
(over $3 \sigma$) from the Paczy\'{n}ski best fit.
We adopt the criterion
{\it (i)} $\chi_{\rm r} > \chi_{\rm r{~\rm th}} = 4$ and
$N_{\rm good}> N_{\rm good~th}=3$.
In other words, the latter condition means that
if we have significant deviations at only one or two points,
we cannot conclude  that they are caused by a planet orbiting the primary lens.

\begin{figure}
\includegraphics[width=80mm]{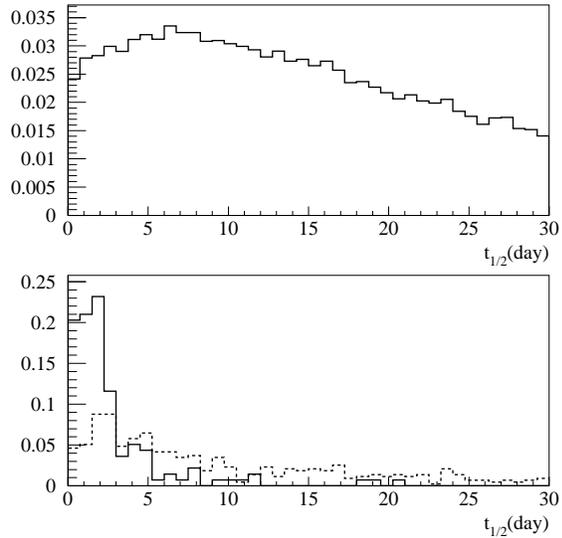}
\caption{
Normalized (to unity) distributions of $t_{1/2}$
(the duration of a microlensing event without
planet). Top panel: detectable events.
Bottom panel: selected events ($\chi_{\rm r} > 4 $, $N_{\rm good}>3$
and $ \langle \epsilon \rangle _{\rm max} >0.1$) for I class ($\rho/u_0>1$, solid line) and
II class ($\rho/u_0<1$,  dashed line) events.
In Figs. 1 -- 11 we take $D=8$ m telescope parameters and $N_{\rm im}=12$
day$^{-1}$.} \label{figura2}
\end{figure}

\begin{figure}
\includegraphics[width=80mm]{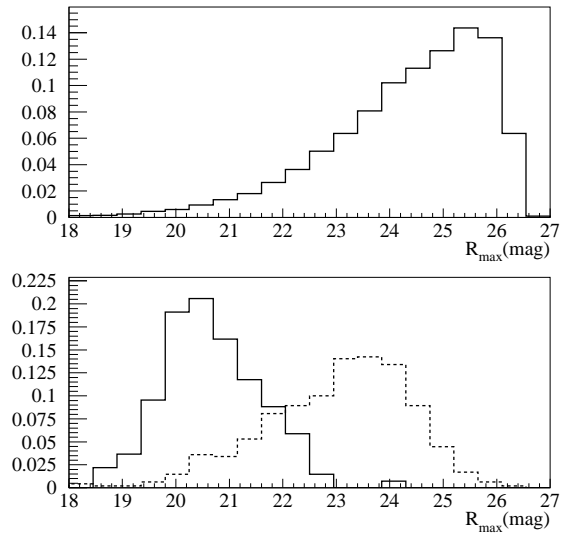}
\caption{Normalized (to unity) distributions of $R_{max}$
(the magnitude corresponding to the flux variation
at the maximal Paczy\'{n}ski  magnification).
Top panel: detectable events.
Bottom panel: selected events for I class ($\rho/u_0>1$, solid line) and
II class ($\rho/u_0<1$, dashed line) events.
Here we take $D=8$ m telescope parameters.} \label{figura3}
\end{figure}

The light curves fulfilling the above condition {\it (i)}
may show only an overall distortion with respect
to the underlying Paczy\'{n}sky shape.
This is characteristics, in particular, for events with
large source radii and small planetary masses.
Our second criterion is therefore meant to look for
and quantify the single more significant
planetary perturbations. To this purpose
we consider the time dependent,
average (with respect to the source area $\Sigma$)
relative planetary  magnification
\begin{equation}
\langle \epsilon(t) \rangle = \left( \frac
{\int_{\Sigma} d^2 \vec{x} ~
[|A_{\rm P}(\vec{x},t)-A^0(\vec{x},t)|/A^0(\vec{x},t)]}
{\int_{\Sigma} d^2 \vec{x}} \right) ~.
\end{equation}
This quantity is sensibly different from zero only when,
depending on the source and lens parameters and relative motion,
there is (at a given time) a substantial overlapping between the source area
and the caustic (central and/or planetary) region.
So, to select light curves with detectable planetary
features, besides condition {\it (i)}, we further require that {\it (ii)}
there exist at least one point on the light curve with
$\langle \epsilon \rangle_{\rm max}$ larger than
$\langle \epsilon \rangle_{\rm th} = 0.1$.
By using both conditions the number of selected events get reduced
of about 50\% with respect to the events selected by using
only the condition {\it (i)}.
The condition {\it (i)} is particularly efficient to select
light curves with a large number of points deviating from the
Paczy\'{n}ski fit,
the condition {\it (ii)} ensures the presence on the light curve
of at least one clear planetary feature.
Note that in this analysis we do not attempt to further characterize the
planet deviations as due to intersection of central and/or planetary caustics.

\section[]{Results}

\begin{table*}
\centering
\caption{Parameters of events shown in Figs. \ref{1428} - \ref{500}.
We also give some microlensing parameters and
in the last three  columns the sum of residuals $\chi_{\rm r}$
along the whole light curve, the sum of residuals $\chi_{\rm r~max}$
and the maximum value of the relative planetary magnification
$\langle \epsilon \rangle_{\rm max}$ during the time interval corresponding
to the strongest planetary feature.}
\medskip
\begin{tabular}{|c|c|c|c|c|c|c|c|c|c|c|c|c|}
\hline
\hline
   &$\rho/u_0$&   $u_0$                &$d_{\rm P}/R_{\rm E}$ &$M_{\rm P}$         &$\theta$& $R_{\rm E}$ &  $t_{\rm E}$&$R_{\rm max}$& $t_{1/2}$& $\chi_{\rm r}$ & $\chi_{\rm r~max}$ & $\langle\epsilon\rangle_{\rm max}$ \\
   &          &                        &          &($M_{\rm{Jupiter}})$& (deg)  &  (AU) & (day) & (mag)   & (day)    &          &              &       \\
\hline
\#1 &   2.89   &  $9.47 \times 10^{-3}$ &  0.90    &  4.74       & 341.8   &   2.2 &  16.1 &  20.2 &   0.5     & 194      & 730          & 0.64 \\
\#2 &   1.18   &  $2.63 \times 10^{-2}$ &  0.68    &  0.82       & 104.8   &   2.8 &  52.1 &  21.1 &   4.0     & 43       & 980          & 0.25 \\
\#3 &   0.12   &  $3.56 \times 10^{-1}$ &  2.25    &  0.22       & 190.3   &   2.3 &  18.7 &  23.5 &  16.5     & 13       &  79          & 0.57 \\
\#4 &   0.08   &  $1.62 \times 10^{-1}$ &  1.32    &  3.97       & 336.0   &   3.9 &  28.4 &  23.8 &  13.3     & 37       & 153          & 1.13 \\
\hline
\hline
\end{tabular}
\label{tab2}
\end{table*}


\begin{figure}
\includegraphics[width=80mm]{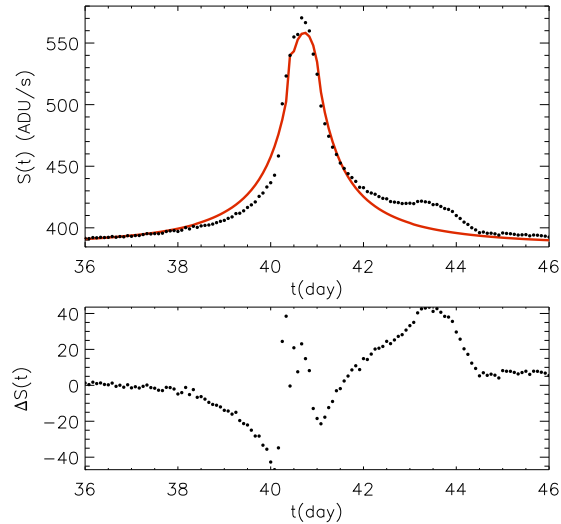}
\caption{
Event of the I class \#1 (see Table \ref{tab2}).
The upper panel shows the simulated light curve (black dots)
and the corresponding best fit model (continuous line), that is
a Paczy\'{n}ski light curve modified for finite source effects.
The bottom panel gives the difference.}
\label{1428}
\end{figure}

\begin{figure}
\includegraphics[width=80mm]{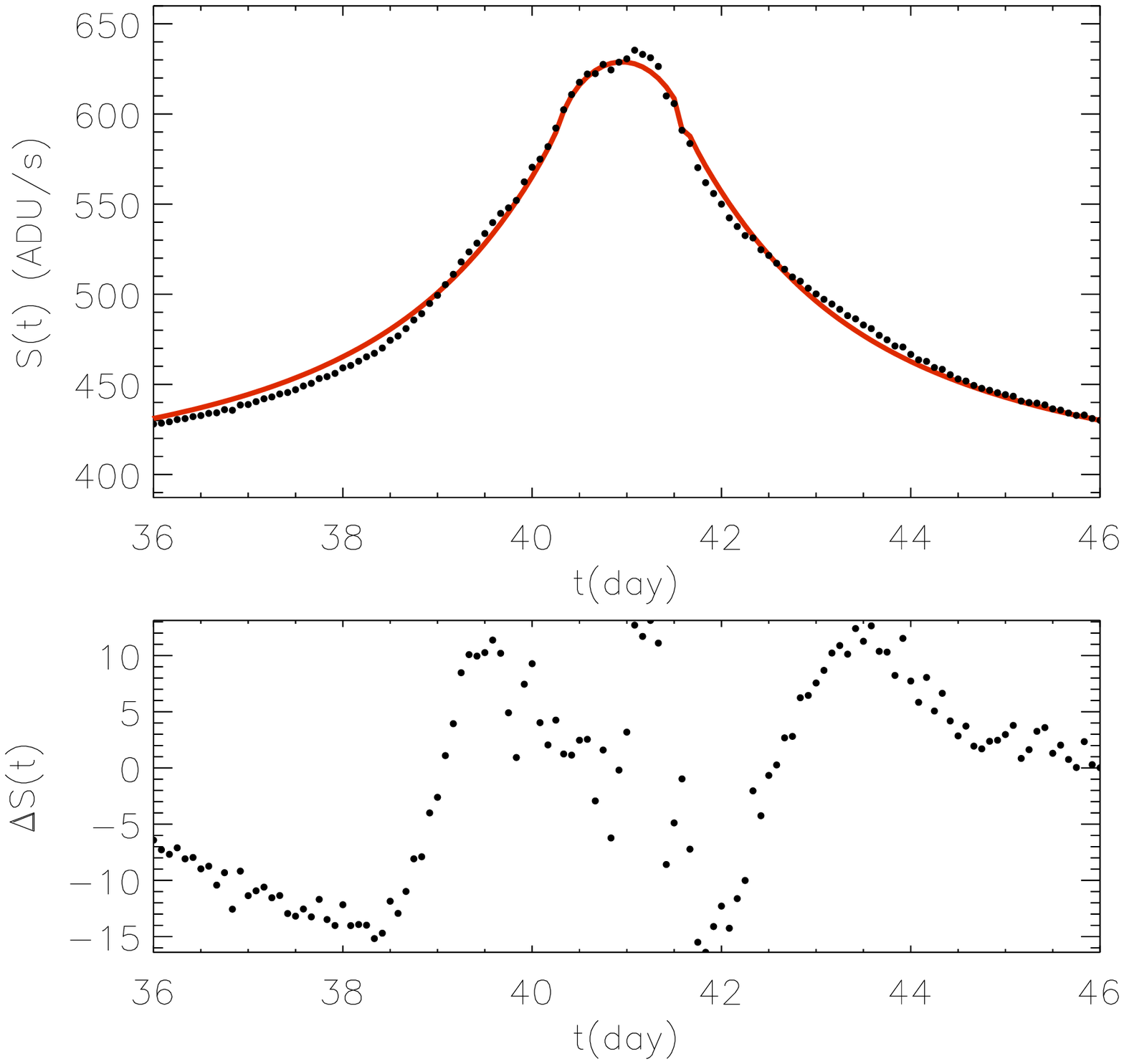}
\caption{The same as in Fig. \ref{1428} for the
I class event \#2 (see Table \ref{tab2}).}
\label{1083}
\end{figure}

\begin{figure}
\includegraphics[width=80mm]{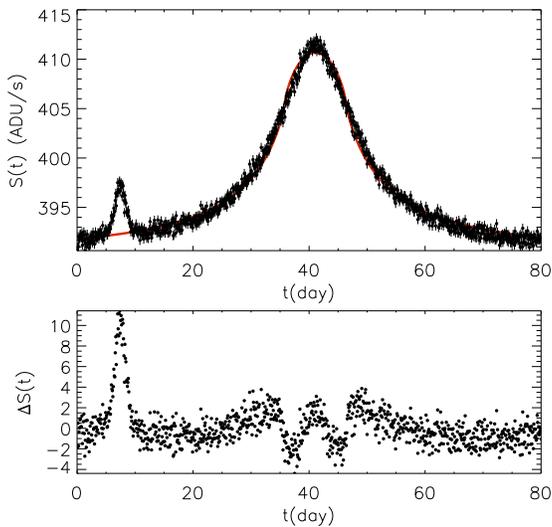}
\caption{The same as in Fig. \ref{1428} for the
II class event \#3 (see Table \ref{tab2}).
We note that the fit follows the simulated data except for a
small time interval ($5<t<10$ day).}
\label{3056}
\end{figure}

\begin{figure}
\includegraphics[width=80mm]{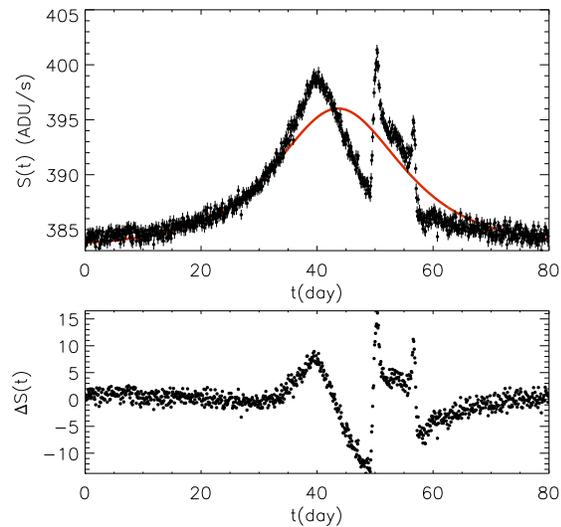}
\caption{The same as in Fig. \ref{1428} for the
II class event \#4 (see Table \ref{tab2}).}
\label{500}
\end{figure}

In the following analysis we consider four different telescope diameters
$D=1.5,~2.5,~4$ and 8 m (corresponding to zero-point in the $R$-band of
$23.1~,24.3,~25.3$ and 26.8 mag, respectively), $t_{\rm exp}=30$ min for
all cases and we take $N_{\rm im} = 12 $
day$^{-1}$, corresponding to a regular sampling time of two hours.
The effect of taking larger telescopes is that of increasing
the number of faint, detectable events. Moreover,
we consider only self-lensing events towards the four considered lines of
sight (see Section 3.1), leaving out the eventual MACHO component in the
galactic halos and assume the presence of a planet orbiting around
each star in the M31 bulge and disk.

An advantage of the Monte Carlo approach to the binary microlensing analysis
is that we can characterize the events with planetary detections.
We remind that these events have been selected,
from the whole sample of detectable events,
by requiring
$\chi_{\rm r} > 4$, $N_{\rm good} > 3$ and
$\langle\epsilon\rangle_{\rm max}>0.1$.

In the Figs. \ref{figura2} and \ref{figura3} (for $D=8$ m)
we give the distributions of $t_{1/2}$ and $R_{\rm max}$ for detectable
(top panels) and selected events (bottom panels)
\footnote{We notice that the distributions
of $t_{1/2}$ and $R_{\rm max}$ for detectable and selected events
weakly depend on telescope diameter $D$.}.
As usual (see, e.g., \citealt{kerins01}),
$t_{1/2}$ is the full-width half-maximum microlensing event duration
and $R_{\rm max}$ the magnitude in the $R$-band corresponding
to the flux variation at the maximal Paczy\'{n}ski  magnification.
Comparing the corresponding distributions,
we see that events with short time duration
and large flux variation (therefore with smaller impact parameter)
have a larger probability to show planetary deviations.
This result is due to the fact that the crossing of the central caustic
(close to the primary lens star) by the source trajectory
is more probable in events with source and lens closely aligned.

As next, for the selected events (bottom panels in the Figs. \ref{figura2}
and \ref{figura3}) we discriminate two classes of events
(indicated by I and II), according to the ratio
$\rho/u_0 >1$ (solid lines), or, $\rho/u_0 <1$ (dashed lines).
The ratio $\rho/u_0$  characterizes the relative size
of the source with respect to the event geometry, since
$\rho$ is the dimensionless source radius
in the lens plane and $u_0$ is the dimensionless lens
impact parameter.
The I class of events with $\rho/u_0>1$  is accounted for events with shorter
time duration and higher  magnification  at maximum. The median values of the
two distributions are $(t_{1/2})_{\rm med}=1.6$ day and
$(R_{\rm max})_{\rm med}=20.6$ mag.
Two light curves of I class events are shown in the
Figs. \ref{1428} and \ref{1083}.
The first figure (for $M_{\rm P} = 4.74~M_{\rm {Jupiter}}$)
show a more clear deviation with respect to the Paczy\'{n}ski light curve.
The second one (for $M_{\rm P} = 0.82~M_{\rm {Jupiter}}$),
which is representative
from a statistical point of view of the
whole sample of I class events,
shows an overall distortion (that in other cases may be either symmetric
or asymmetric with respect to the maximum) of the light curve.
\begin{figure}
\includegraphics[width=80mm]{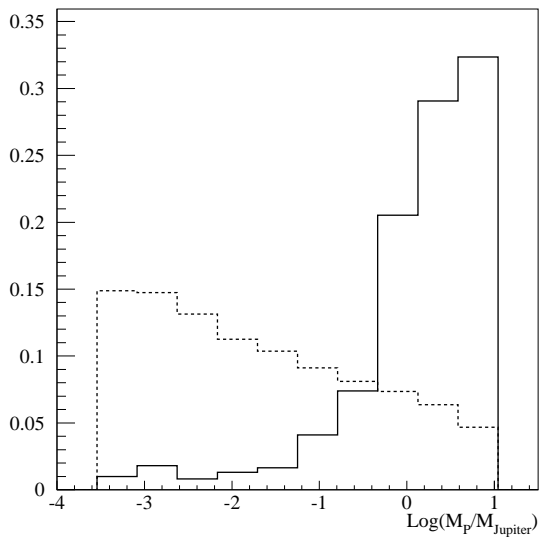}
\caption{Normalized (to unity) distributions
of the planet mass $M_{\rm P}$ for the
events with detectable planetary deviations (solid line)
and for the generated events (dashed line).
Here we take $N_{\rm im}=12$ day$^{-1}$ and $D=8$ m.}
\label{figura4}
\end{figure}
\begin{figure}
\includegraphics[width=80mm]{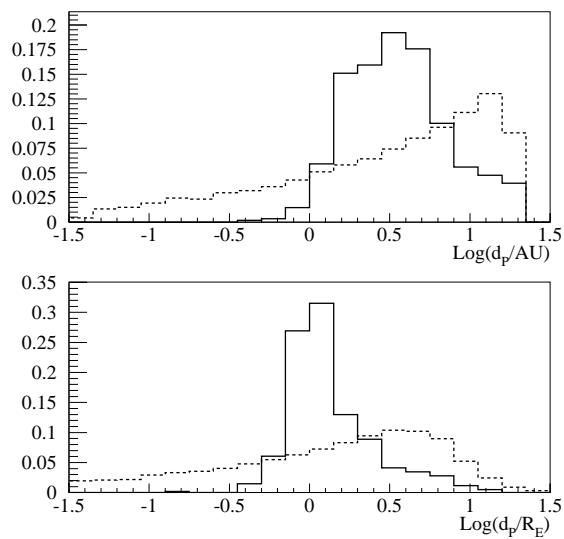}
\caption{Upper panel: (normalized to unity)
distributions of the star-to-planet separation
$d_{\rm P}$ (in AU units) for the events with detectable planetary deviations
(solid line) and for the generated events (dashed line).
Bottom panel: distributions of $d=d_{\rm P}/R_{\mathrm E}$ for events as
before.}
\label{figura5}
\end{figure}
As far as the II class of events with $\rho/u_0<1$ is concerned,
the dashed lines in the bottom panels of the
Figs. \ref{figura2} and \ref{figura3}
show that they have larger time duration - $(t_{1/2})_{\rm med} = 6.4$ day -
and lower  magnification  at the
maximum - ($R_{\rm max})_{\rm med}=23.1$ mag -.
Two examples of light curves are given in
Fig. \ref{3056} (for $M_{\rm P} = 0.22~M_{\rm {Jupiter}}$)
and  in Fig. \ref{500} (for $M_{\rm P}=3.97~M_{\rm {Jupiter}}$),
with a bump and a multiple-peak structure,
which is typical of binary microlensing
(in which the companion mass is large).
These features of caustic intersections were discussed also
by \cite{Paczynsky_96}.

Concerning the reliability of the planetary detections,
we find that the events of the I class (with $\rho/u_0>1$)
have smaller values of
$\langle \epsilon \rangle_{\rm max}$ (for a given $M_{\rm P}$ value)
with respect to the II class events. This happens since for the I class
events the source size $\rho$ is typically much larger than the caustic
region, so that averaging the planetary  magnification  on the source area
leads to smaller values of $\langle\epsilon\rangle_{\rm max}$.
This does not occurs for the events of the II class (with $\rho/u_0<1$),
for which averaging on the source area is less important.
This result is reflected in the presence of more clear and temporally
localized planetary features in the II class events. These deviations
look similar to that observed in microlensing planetary events
towards the galactic bulge, for which the point-like source approximation
is acceptable. We also find that
$\langle \epsilon \rangle_{\rm max}$ increases with
increasing values of $M_{\rm P}$,
a result that is expected since the caustic size
is increasing.

\begin{figure}
\includegraphics[width=80mm]{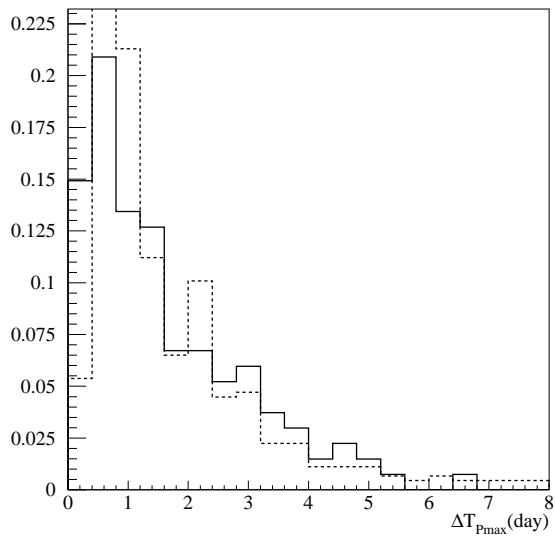}
\caption{Normalized (to unity) time duration $\Delta T_{\rm P~max}$
distribution of the strongest planet induced perturbation,
for I class (solid line) and II class (dashed line) events.}
\label{figura6}
\end{figure}

\begin{figure}
\includegraphics[width=80mm]{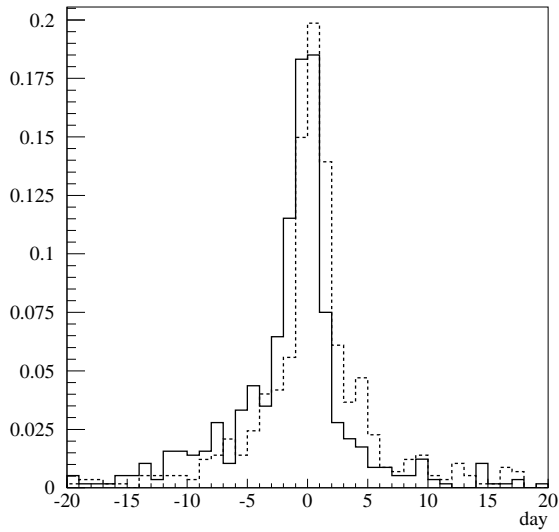}
\caption{
Histograms for the differences $(t_{\rm i}-t_0)$ (solid line) and
$(t_{\rm f}-t_0)$ (dashed line) for
the selected events
($\chi_{\rm r} > 4 $, $N_{\rm good}>3$ and
$ \langle \epsilon \rangle_{\rm max} > 0.1$).
Initial and final instants for the start and the end of the
strongest deviation in the light curves
are denoted by $t_{\rm i}$ and $t_{\rm f}$, while $t_0$
is the instant of maximum on the light curve.}
\label{figura9}
\end{figure}

The distributions of the planet mass $M_{\rm P}$
(for $D=8$ m and the considered lines of sight)
are given in the Fig. \ref{figura4} (solid line) for the selected events.
($\chi_r > 4$, $N_{good} >3$ and $\langle\epsilon\rangle_{rm max} > 0.1$).
For comparison, the $M_{\rm P}$ distribution for the whole sample of detectable
events (dashed line) is also given.
From Fig. \ref{figura4} it follows that larger planetary masses lead to
higher probability for the detection of planetary features.
This result reflects the fact that the detection probability is
proportional to the caustic size, which increases with the
planet-to-star mass ratio \citep{maopacz91,bolatto,gould92}.
From the same figure, it also follows that the planet detection
can occur with a non negligible probability
for $M_{\rm P} > 0.06~M_{\rm {Jupiter}}$ ($M_{\rm P} > 20~M_{\oplus}$),
although even Earth mass planets might be in principle detectable.
However, if we consider telescopes with smaller diameter,
practically no planet detection
occurs for $M_{\rm P} < 0.06~M_{\rm {Jupiter}}$ and $D <4 $ m.

We also recover the well known result that the probability of planet
detection is maximized when the planet-to-star separation $d_{\rm P}$
is inside the
``lensing zone'' \citep{gould92,griest98}. The $d_{\rm P}$
(normalized to unity) distribution
for selected (solid line) and detectable (dashed line) events are shown in
the upper panel of Fig.~\ref{figura5}.
The relevance of the lensing zone is clarified in the
bottom panel of the same figure where the planet separation (in unit of the
Einstein radius) $d=d_{\rm P}/R_{\mathrm E}$  is plotted.
It turns out that about 70\% of events with planet detections
have $d$ values distributed in the lensing zone.
We also find an excess of I class events at large planetary distances $d>1.6$,
that is related to the interplay between the source size and the size of the
central caustic.

\begin{table*}
\centering
\caption{Pixel-lensing events with positive planetary detections
($\chi_{\rm r} > 4$, $N_{\rm good}> 3$ and
$\langle\epsilon\rangle_{\rm max}>0.1$).
Median values of the considered distributions.
Upper part of the table: I class events ($\rho/u_0 > 1$).
Lower part: II class events ($\rho/u_0 < 1$).}
\medskip
\begin{tabular}{|c|c|c|c|c|c|c|}
\hline
\hline
              & $(R_{\rm max})_{\rm med}$ & $(t_{1/2})_{\rm med}$ & $(d_{\rm P})_{\rm med}$ & $(M_{\rm P})_{\rm med}$         &  $(\Delta {\rm T~max})_{\rm med}$ & $(\Delta T_{\rm P~tot})_{\rm med}$ \\
              &   (mag)           &   (day)           & (AU)          & $(M_{\rm {Jupiter}})$ & (day)                   & (day)                  \\
\hline
\hline
I class       &   20.6            &    1.6            &   4.5         & 1.56                  &        1.5              &         3.4             \\
 $\rho/u_0>1$ &                   &                   &               &                       &                         &                        \\
\hline
II class      &   23.1            &    6.4            &   3.3         & 2.09                  &        1.2              &         1.6              \\
 $\rho/u_0<1$ &                   &                   &               &                       &                         &                        \\
\hline
\hline
\end{tabular}
\label{tab3}
\end{table*}
The knowledge of the typical time scales for the planetary perturbations
is an important issue to choose an adequate strategy for the observations,
namely, telescope parameters and suitable sampling time for optimizing
the detection of the planetary perturbations in the light curves.
To estimate the time duration of the strongest perturbations
we introduce a new estimator, $\chi_{\rm r~n}$, that is defined as
$\chi_{\rm r}$ with the difference that now
the sum runs over the points inside the
$n$-th planetary perturbation.
We consider a perturbation to be significative
whenever $\chi_{\rm r~n}>4$.
The duration $\Delta T_{\rm P~max}$
is estimated as the time interval
with the largest value of $\chi_{\rm r~n}$.
The normalized distribution of $\Delta T_{\rm P~max}$
is shown in Fig.~\ref{figura6}. It results that in pixel-lensing searches
towards M31 typical time duration of the strongest planetary perturbations
is about 1.5 and 1.2 days for I and II class events, respectively.
The normalized distributions of the initial ($t_{\rm i}$) and
final ($t_{\rm f}$) instants for the start and the
end of the strongest planetary deviations
($\Delta T_{\rm P~max} = t_{\rm f}-t_{\rm i}$) given in Fig. \ref{figura9}
show that these occur near the maximum  magnification time, as expected since
in pixel-lensing the crossing of the central caustic is more probable.
We also find that the number of time intervals with significative
planetary deviations on each light curve increases with
increasing values of the ratio $\rho/u_0$.
Indeed, the overall time scale
$\Delta T_{\rm P~tot} = \sum_n \Delta T_{\rm P~n}$
for the significative planetary deviations increases up to 3.4 and 1.6 days
for I and II class events, respectively.
Moreover, our analysis of the distribution of $\Delta T_{\rm P~tot}$
as a function of telescope size $D$ and sampling time
$N_{\rm im}^{~-1}$ allows us to conclude
that a reasonable value of the time step
for pixel-lensing observations aiming to detect planets in M31 is a few hours
($N_{\rm im} \simeq 4$ day$^{-1}$), almost irrespectively on $D$.

To summarize, the distinctive features of the selected events with planetary
detections are given by in Table \ref{tab3}
(for a telescope with $D=8$ m and averaging
on the considered lines of sight). In particular,
we report the median values for the
distributions of the more relevant quantities characterizing the lensing
and planetary systems.

\begin{table}
\centering
\caption{As a function of $D$ (first column) we give:
the probability to detect pixel-lensing events (second column)
normalized to the events detectable by a 8m telescope,
the fraction of I class (third column) and
               II class (fourth column) events,
the probability to detect planetary features
($\chi_{\rm r} > 4$, $N_{\rm good}> 3$ and
$\langle\epsilon\rangle_{\rm max} > 0.1$)
for I (fifth column) and II (sixth column) class of events
when normalized to the events detectable by a telescope with diameter $D$
and the overall probability (last column).
Here we assume $N_{\rm im}=12$ day$^{-1}$ and $t_{\rm exp} =30$ min.}
\medskip
\begin{tabular}{|c|c|c||c|c|c|c|}
\hline
\hline
     $ D$   &  $\Gamma(D)/\Gamma(8)$& $f^{I}$ &  $f^{II}$      &   $P_{\rm P}^I$   &   $P_{\rm P}^{II}$&  $P_{\rm P}$    \\
     (m)    &          (\%)         &         &                &      (\%)   &    (\%)     &    (\%)   \\
\hline
   1.5&    27        &   0.15              &  0.85           &               0.8   &     0.1  & 0.2    \\
   2.5&    62        &   0.07              &  0.93           &               2.8   &     0.4  & 0.6    \\
   4  &    78        &   0.06              &  0.94           &               4.8   &     0.8  & 1.1    \\
   8  &    100       &   0.04              &  0.96           &               9     &     1.2  & 1.5    \\
\hline
\hline
\end{tabular}
\label{tab1}
\end{table}

In Table \ref{tab1}  we present the planet detection probabilities
(by averaging on the selected lines of sight),
assuming $N_{\rm im}=12$ day$^{-1}$, $t_{\rm exp}=30$ min and
telescopes of different diameters
\footnote{
Note that, since in pixel-lensing the important parameter is the
signal-to-noise ratio and it is proportional to
$D \sqrt{t_{\rm exp}}$, to have the same probability for
planetary feature detection, one can use smaller size telescopes as well,
by increasing correspondingly the exposure time.}.
For each telescope
diameter and class of events, the probabilities are evaluated
as the ratios between the number of the selected events
and the number of events detectable for the same class and telescope, namely
$ P_{\rm P}^I = \Gamma^I_{\rm P}(D)/\Gamma^I(D)$ and
$ P_{\rm P}^{II} = \Gamma^{II}_{\rm P}(D)/\Gamma^{II}(D)$.
The fractions $f^I$ and $f^{II}$  of detectable events for each class
is also given in Table \ref{tab1}.
It results that the probability to detect planetary signatures
is higher for the events of the I class (with $\rho/u_0>1$),
that however are rare.
On the contrary, the generated events of the  II class are more
numerous, but have a smaller probability to show
detectable planetary features.
The overall probability ($P_{\rm P}$
in the last column of the Table \ref{tab1})
is always very small (less than 2~\%) and decreases
rapidly for smaller telescopes.
This implies that hundreds of pixel-lensing events should be collected
to find a few systems with planetary features.
\begin{figure}
\includegraphics[width=80mm]{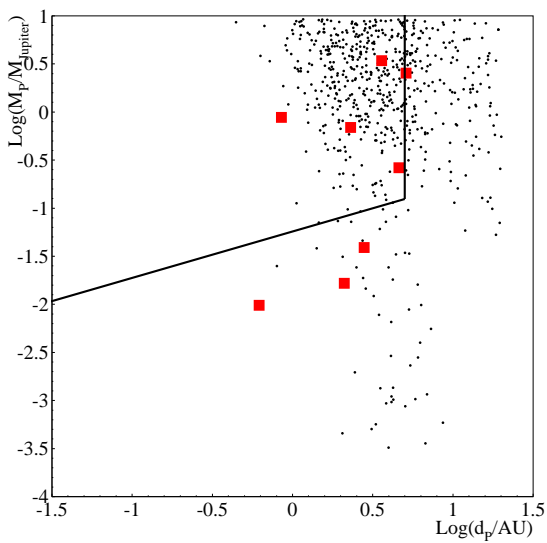}
\caption{
Scatter plot of the planet mass (in unit of Earth mass) vs
planet distance (in Astronomical Units).
The solid thick line delimits the region (upper and left) of planet
detection accessible by radial velocities measurements with a precision up to
1 m s$^{-1}$. The observational data were accessed using the extrasolar
planet on-line catalogue which collects the results of several collaborations
(see http://exoplanet.eu/catalog.php and references therein).
The eight small boxes are the planets detected by the microlensing technique.
Starting from a sample of 40,000 detectable  pixel-lensing events ($D=8$ m),
630 selected events (indicated by black dots)
with $\chi_{\rm r} > 4$, $N_{\rm good} >3$ and $\langle\epsilon\rangle_{\rm max} > 0.1$
show planetary features and among these 48 events have
$M_{\rm P}<20~M_{\oplus}$.}
\label{figura7}
\end{figure}
\begin{figure}
\includegraphics[width=80mm]{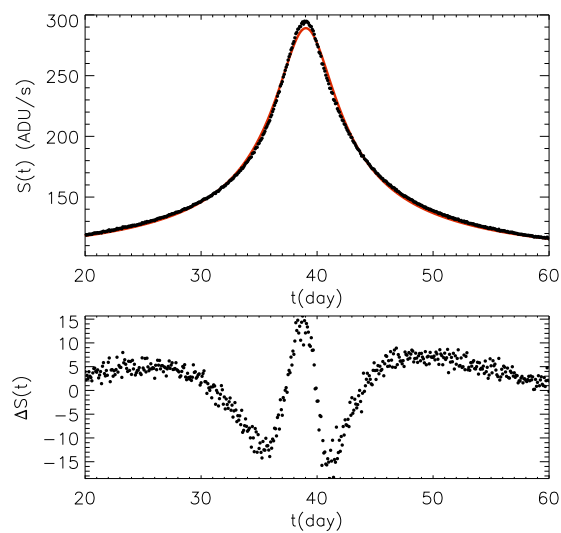}
\caption{
The upper panel shows the simulated light curve (black dots)
of a planetary event with the
parameters of the best fit model for the PA-99-N2 event
(model W1 in Table 1 of \citealt{an04}).
In particular, $d=1.84$, $q = 1.22 \times 10^{-2}$
(corresponding to a planet mass $M_{\rm P} = 6.34~M_{\rm {Jupiter}}$
 for a disk lens of mass $M_l = 0.5~M_{\odot}$),
$u_0 = 3.4 \times 10^{-2}$, $t_{\rm E} = 132.3$ day
and $\theta = 24.5$ deg.
We take the source magnitude $M_{\rm R} =-2.0$, and the source radius of
$R_s = 11~R_{\odot}$ (corresponding to $\rho = 1.27 \times 10^{-2}$ and
$\rho/u_0 = 0.37$).
It is also shown the best fit Paczy\'{n}ski like model modified for
finite source effects (continuous line), which appears
almost indistinguishable from the simulated data.
The bottom panel gives the difference between the two curves.
Here we use the INT telescope parameters
and $N_{\rm im} = 12$ day$^{-1}$.}
\label{en2}
\end{figure}

The main result of the present work can be summarized
in Fig.~\ref{figura7} ($D=8$ m), where for self-lensing events towards M31
with detectable planetary features we present the event scatter plot in the
$(M_{\rm P},d_{\rm P})$ plane.
The thick solid lines delimit the region (upper and left)
where extrasolar planets are detectable by ground based observations,
that are more sensitive to massive and close-in planets
and that can be successfully applied only for systems close enough
to Earth.
We remind that current space based observations
by Kepler\footnote{http://www.nasa.gov/mission\_pages/kepler/overview/index.html}
and COROT\footnote{http://smsc.cnes.fr/COROT/index.html} satellites)
are expected to decrease the minimum detectable
planetary mass limit (up to one tenth of the Earth mass) and
increase the planetary distance (up to tens of AUs).
The eight extrasolar planets claimed so far to be detected by microlensing
since 2003 in observations towards the Galactic bulge are represented by boxes.
The locations of points in Fig. \ref{figura7} show that the pixel-lensing
technique may be used to search for extrasolar planets in M31
(including small mass planets), and at the moment this is the only
method to discover planets in other galaxies.
As one can see, detectable extrasolar planets have
planet-to-star separations in the range $0.3 - 25$ AU and
mass in the range $ 0.1~M_{\oplus} - 10 ~M_{\rm {Jupiter}}$
(that coincides with the assumed lower and upper
limits for planetary masses in the simulations). However, we note that
the detection of planets with relative large masses is favourite
(see also Fig. \ref{figura4}).
We also caution that the planets with $M_{\rm P} < 20~M_{\oplus}$
become undetectable and disappear from  Fig. \ref{figura7} if
the adopted telescope has not a good enough photometric stability
(about 0.03 mag, that is the required stability consistent with the typical
error bars for the detection of small mass planets).

Before closing this section we note that an extrasolar planet
in M31 might have been already detected since an anomaly
in a pixel-lensing light curve has been reported \citep{an04}.
The authors claim that a binary system (lying on the M31 disk)
with mass ratio $q = 1.22 \times 10^{-2}$ and distance $d = 1.84$,
is a possible explanation of the anomaly in the observed light curve.
Other parameters are indicated in the caption of Fig. \ref{en2}.
In this figure we give a light curve with the best fit parameters of the
PA-99-N2 event as given in Table 1 of  \cite{an04}.
It gives a clear deviation
($\chi_{\rm r} = 49$, $\langle\epsilon\rangle_{\rm max} = 0.6$)
with respect to the corresponding Paczy\'{n}ski shape,
at least with our ideal sampling of $N_{\rm im} =12$ day$^{-1}$
and observational conditions.
In order to estimate the secondary object mass,
we assume that the disk star mass follows the broken power law
given by \cite{an04}. Accordingly, one finds a mean mass of
$\simeq 0.5~M_{\odot}$ for the lens and therefore a
mean value of $M_{\rm P} = 6.34~M_{\rm {Jupiter}}$ for the planet.
This value
is at the boundary between the planet and brown dwarf region.
Our light curve closely resembles
the observed one and the basic characteristics of the planetary event fall
in the parameter range for the II class of events.

\section[]{Conclusions}

We consider the possibility to detect planets in M31 by using
pixel-lensing observations with telescopes of different
sizes and observational strategies. This is the only way to detect planets
in other galaxies and acquire information allowing a comparison of
the planetary systems in M31 with respect to those in the Milky Way.
We carry out MC simulations and explore the multi-dimensional space of the
physical parameters of the planetary systems and characterize the sample of
microlensing events for which the planet detections are more likely to be
observed.
We have assumed that each lens star in the M31 bulge and disk hosts one planet,
and used for the planet mass distribution an simplified law,
neglecting any dependence of the planet mass on the parent star mass and
metallicity.
Consideration of finite source effects induces a smoothening of the planetary
deviations with respect to the point-like source approximation and, in turn,
decreases the chance to detect planets. It also implies that in pixel-lensing
searches towards M31 only few exposures per day could be enough to detect
planetary features in light curves, at least when using
large enough telescopes.
We find that the pixel-lensing technique favours the detection of large mass
planets ($M_{\rm P} \simeq 2~M_{\rm {Jupiter}}$), even if planets with mass
less than $20~M_{\oplus}$ can be detected (with small probability, however)
by using large telescopes with a sufficient photometric stability.
Microlensing is intrinsically a "no repetition" phenomenon and
variable stars may mimic microlensing events and contaminate the sample
of events attributed to microlensing. Therefore, real
observations should be done at least in two bands,
to check for achromaticity and be confident that the
contamination by variables can be sorted out.
However, a minor chromaticity is expected since
the source limb darkening profile depends on the considered band
and on the spectral type of the source star
(see, e. g., \citealt{BogdanovCherepashchuk_95b,Pejcha}).

Finally, we remark that although we have neglected the contribution to
microlensing
events of MACHOs in both galactic halos (in this respect the estimated
planet rate should be considered as a lower bound),
pixel-lensing observations towards
M31 could be very useful in establishing
whether planets may form around MACHOs as well.

\section*{Acknowledgments}
This work has made use of the IAC-STAR Synthetic CMD computation
code. IAC-STAR is supported and maintained by the computer division
of the Instituto de Astrofisica de Canarias".
SCN acknowledges support for this work by the Italian Space Agency (ASI)
and by FARB-2008 of the Salerno University.
We would like to thank the anonymous referee for his helpful comments.


\end{document}